\documentclass[preprint,aps,prl,showpacs,a4paper,unsortedaddress,amsmath,amssymb,byrevtex]{revtex4}
\usepackage[latin1]{inputenc}
\usepackage{graphicx}
\usepackage{dcolumn}
\usepackage{bm}
\usepackage{hyperref}
\usepackage{times}

\begin{document}
\preprint{ffuov/02-01}

\title{Carbon nanotube electron windmills: a novel design for
nano-motors.}
\author{S. W. D. Bailey}
\author{I. Amanatidis}
\author{C. J. Lambert}
\affiliation{Department of Physics, Lancaster University, Lancaster,
LA1 4YB, U. K.}

\date{\today}

\begin{abstract}
We propose a new drive mechanism for carbon nanotube (CNT) motors,
based upon the torque generated by a flux of electrons passing
through a chiral nanotube. The structure of interest comprises a
double-walled CNT, formed from, for example, an achiral outer tube
encompassing a chiral inner tube. Through a detailed analysis of
electrons passing through such a "windmill", we find that the
current due to a potential difference applied to the outer CNT
generates sufficient torque to overcome the static and dynamic
frictional forces that exists between the inner and outer walls,
thereby causing the inner tube to rotate.
\end{abstract}


\maketitle

The evolution from micro- to nano-electronics, as exemplified by
the exponential growth in mobile communications and personal
computing is paralleled by the more-recent miniaturisation of
mechanical devices, which are currently undergoing a transition
from commercially-available  microelectromechanical structures
(MEMS) to nanometre-scale nanoelectromechanical structures (NEMS).
Whereas microfabricated motors, actuators and oscillators
\cite{Trimmer1997} are typically manufactured by conventional
semiconductor processing techniques, their nanoscale counterparts
are more difficult to realise.

An early example of an artificial NEMS \cite{Cummings2000},
\cite{Ruoff2000} is based on a telescoping structure formed from
multiwalled carbon nanotubes (MWCNTs). These structures possess
novel electrical properties \cite{grace} and extremely low
inter-shell friction \cite{Kolmogorov2000}. The latter discovery
led to the idea of CNT nanomechanical
oscillators\cite{Kolmogorov2000}, \cite{Legoas2002},
\cite{Zheng2002}, \cite{Forro2000}, \cite{Williams2002} with
gigahertz operation frequencies, which is beyond the reach of MEMS
oscillators \cite{Rivera2003}. The ultra-low inter-shell friction
in MWCNTs also underpins recently-developed CNT-based nanomotors
\cite{Fennimore2003}, \cite{Bourlon2004}, which involve a MWCNT
whose outer shell is clamped to two metallic anchor pads and whose
inner shell (or shells) is free to rotate or oscillate. A metallic
plate is deposited on the mobile shell and movement is induced via
an electrostatic interaction between the metallic plate and
external gates.

\begin{figure}
\includegraphics[width=0.75\columnwidth]{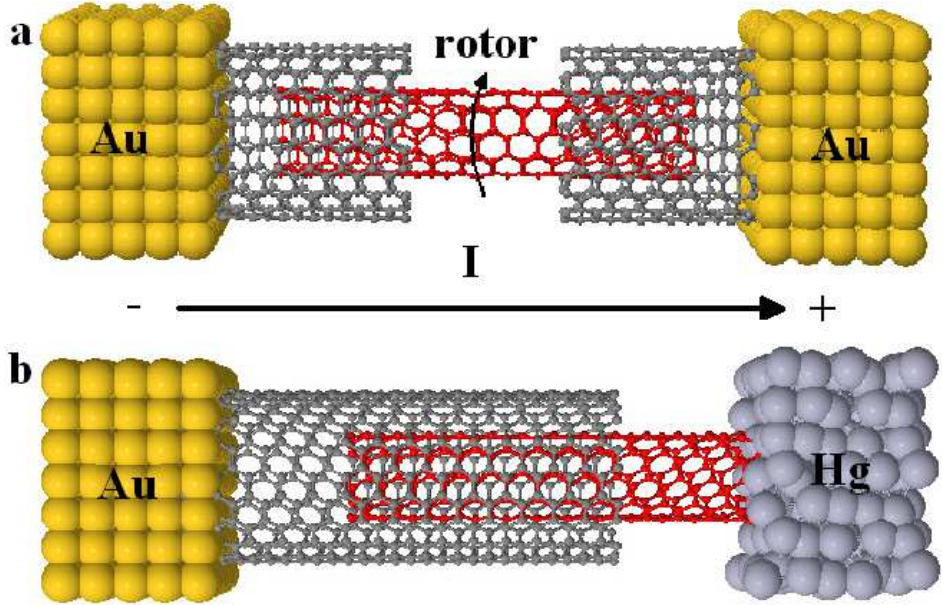}
\vspace{0.2cm} \caption{The proposed nano-motor (1a) and nano-drill
(1b)formed from an inner (6,4) CNT (red) and an outer (18,0) CNT
(grey). The nano-motor is attached to gold electrodes, which act as
reservoirs of electrons, whereas the nano-drill has one end
contacted to a mercury electrode.}
 \label{Fig1}
\end{figure}

The aim of the present paper is to propose a new dc drive
mechanism for CNT-based motors. For all such nano-mechanical
devices analysed to date, the static forces can roughly be
classified as elastic, electrostatic, friction and van der Waals.
In this paper, we propose a new force, which has so far been
ignored in the NEMS literature. This force provides a  new
"electron-turbine" drive mechanism for CNT-based nanomotors and
obviates the need for metallic plates and gates in the nano-motor
of \cite{Fennimore2003}, \cite{Bourlon2004}. To understand the
origin of this force, consider the structure shown in
Fig.~\ref{Fig1}a, which comprises a double-walled CNT, formed from
an achiral (18,0) outer tube clamped to external electrodes and a
chiral (6,4) inner tube. As in \cite{Fennimore2003},
\cite{Bourlon2004}, the central region of the outer tube has been
removed to expose the free-to-rotate, chiral inner tube. The
proposed force arises when a dc voltage is applied between the
external electrodes, which produces a "wind" of electrons (eg)
from left to right. The incident electron flux (from the achiral
CNT) possesses zero angular momentum, whereas after interacting
with the chiral nanotube, the outgoing current carries a finite
angular momentum. By Newton's third law, this flux of angular
momentum produces a tangential force (and an associated torque) on
the inner nanotube, causing it to rotate.

 Fig.~\ref{Fig1}b shows
an even simpler version of this motor, which we refer to as a CNT
"drill", which comprises an achiral outer tube clamped to an
external electrode and a free-to-rotate, chiral inner tube
contacted to a mercury bath, as in \cite{Franck1998}.

\begin{figure}
\includegraphics[width=0.5\columnwidth]{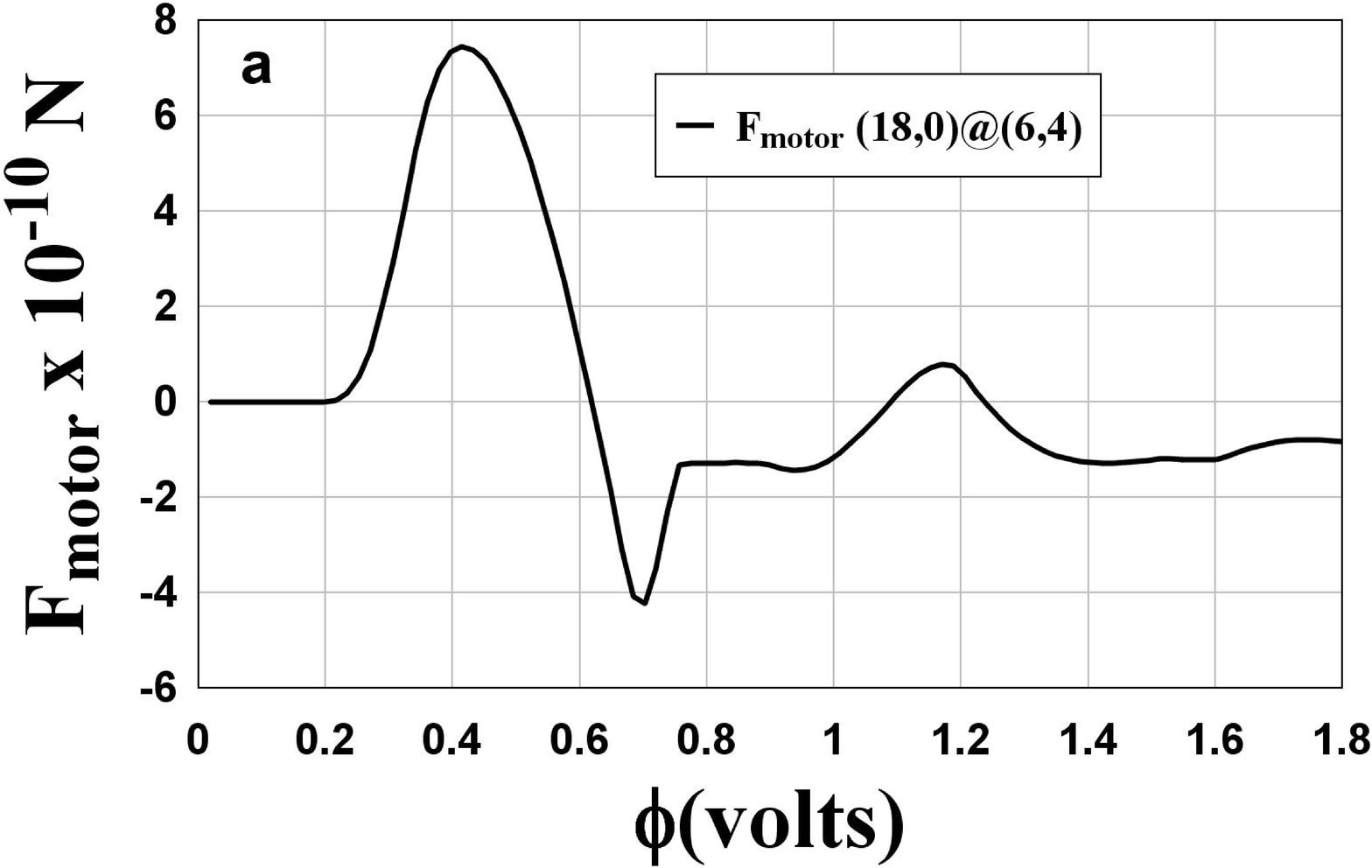}
\includegraphics[width=0.5\columnwidth]{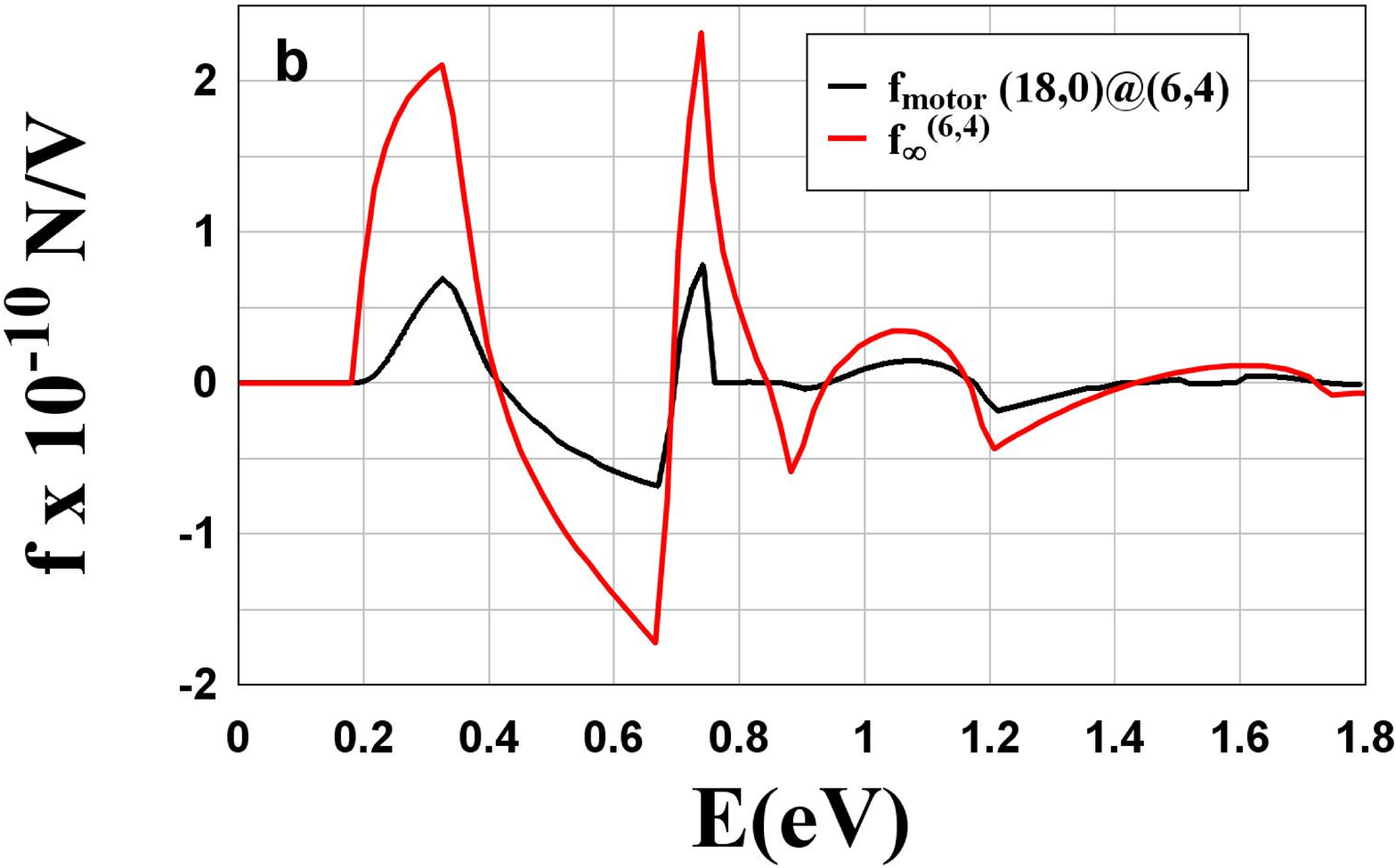}
\includegraphics[width=0.5\columnwidth]{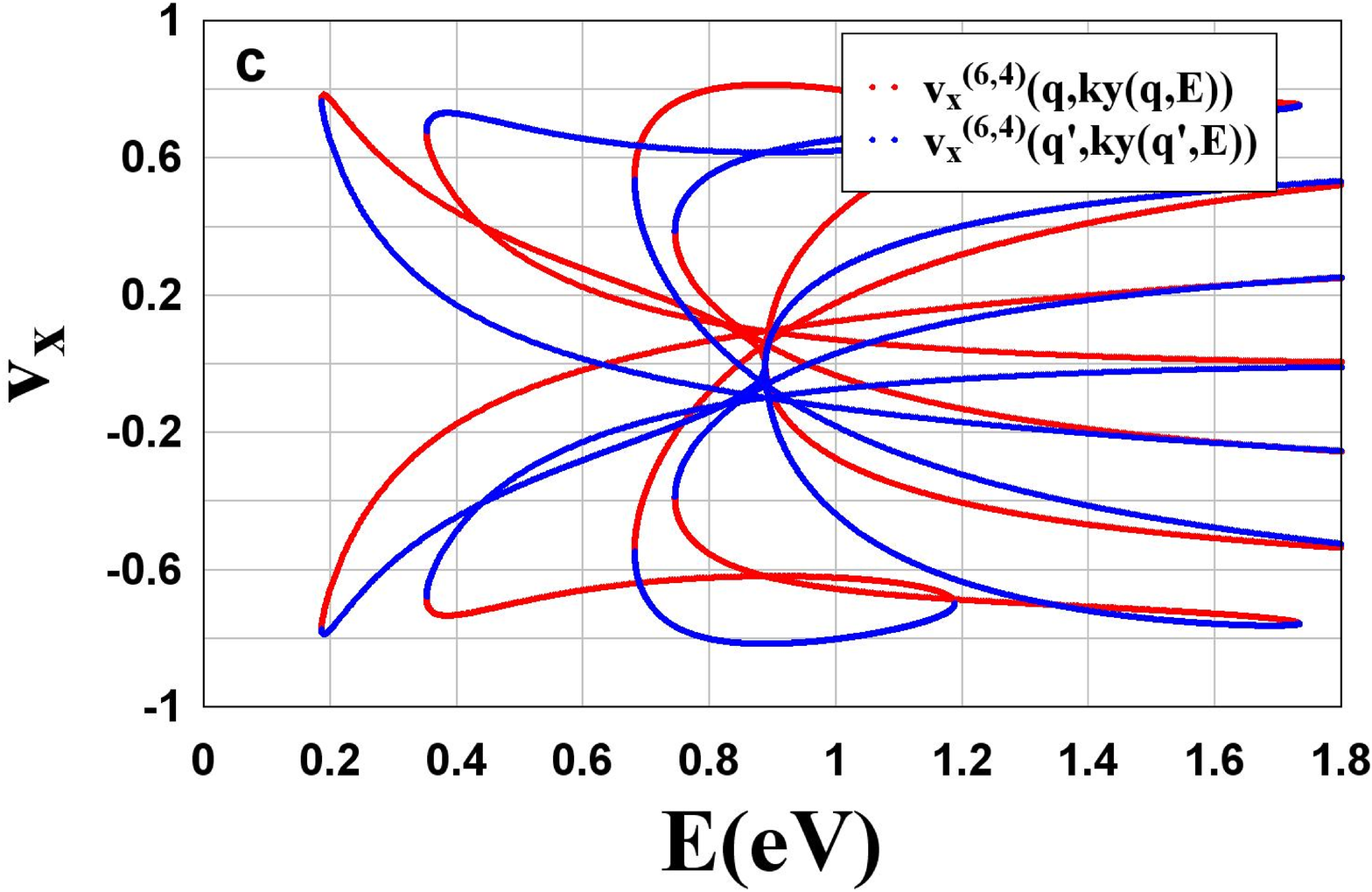}
\vspace{0.2cm}

\caption{$\mathbf{(a)}$ The tangential force (${\rm F_{motor}}$)
exerted on a (18,0)@(6,4) nano-drill as a function of the voltage
$\phi$. $\mathbf{(b)}$. The black curve shows the computed
tangential force per volt ${\rm f}_{{\rm motor}}(E)$ for the
$(18,0)@(6,4)$ drill, whose integral yields the curve in fig 1a,
in accordance with Equ. (7). For comparison, the red curve shows
the quantity ${\rm f}^{(6,4)}_{{\rm \infty}}(E)$ of equation (12),
which is proportional to the total tangential velocity carried by
all right-moving channels of energy $E$ in an infinite (6,4)
chiral CNT. (c) A "fan diagram" showing the tangential velocities
$\mathbf{(c)}$ ${\rm v}_x^{(6,4)}(q, k_y(q,E))/{\rm v_{F}}$
carried by right-moving channels $q$ (red) and left-moving
channels $q'$ (blue) of an infinite (6,4) chiral CNT. The red
curves sum to yield the red curve of fig 2b, in accordance with
Equ. (10).}

\label{Fig2}
\end{figure}

The main question, is whether or not this new force is sufficient
to overcome frictional forces between the inner and outer tube.
The central result of our calculations, is that for moderate
voltages, the tangential force produced by the electron wind can
significantly exceed the frictional forces and therefore electron
windmills provide a viable alternative to electrostatic
nano-motors realised to date. This result is illustrated in
Fig.~\ref{Fig2}a, which shows the tangential force ($F_{\rm
motor}$) as a function of the applied voltage ($\phi$), exerted on
the (18,0)@(6,4) drill of Fig.~\ref{Fig1}b. In this example, the
number of overlap atoms ($N_{atoms}$) between the inner (6,4) and
the outer (18,0) CNT is approximately 4000 and since the static
friction is $F_{\rm exp}\approx 10^{-15}$ N/atom \cite{Kis2006},
the total static friction is $F_c \approx 4.10^{-12}$.
Fig.~\ref{Fig2}a shows that when $\phi$ is approximately 0.4
volts, $F_{\rm motor}$ exceeds $F_c$ by almost three orders of
magnitude.

Before describing the theory leading to Fig.~\ref{Fig2}, it is
also of interest to ask if $F_{\rm motor}$ can overcome the
dynamical friction between the two CNTs \cite{Zhang2004},
\cite{Servantie2006}, \cite{Gaspard2006}. For a spinning CNT
 with carbon atoms of mass $m_c$ moving with a tangential velocity
 $v_c$, the dynamical friction force per carbon atom is \cite{Gaspard2006}
$m_c v_c/\tau$, where $\tau \approx 0.5 ns$. Equating the
dynamical friction to $F_{\rm motor}$ yields a maximum tangential
velocity of $v_{\rm max} \approx 10^7 \alpha/N_c ms^{-1}$, where
$N_c$ is the number of overlap atoms and $\alpha = F_{\rm
motor}/F_0$, where $F_0=4.1 \, 10^{-10} N$ is a natural unit of
force (see below). This means that even for moderate voltages, the
electron wind is sufficient to drive the inner tube to the
mechanical breakdown velocity ($\approx 8.10^3 ms^{-1}$)
\cite{Gaspard2006}.

This demonstration that an "electron wind" can cause a chiral "CNT
windmill" to rotate, could lead to range of applications
\cite{lamb}. By analogy with the occurrence of spin torques in
magnetic point contacts and tunnel
junctions\cite{Kiselev2003},\cite{Fuchs2006}, which have
applications in nanoscale magnetic memory devices, one expects CNT
windmills to have nanoscale memory applications. For example, a
voltage pulse through a CNT motor would cause the inner element to
rotate by a pre-determined angle, which may be utilized in a
switch or memory element. A rotating chiral CNT in contact with a
reservoir of atoms or molecules could act as a nano-fluidic pump.

We now present details of the calculations leading to Fig 2a. To
calculate the rotational driving force, we note that the momentum
of an electron wavepacket incident along a semi-infinite CNT is
equal to the mass of the electron multiplied by its group
velocity. To quantify the tangential momentum, consider an $(n,m)$
CNT defined by a chiral vector $\mathbf{C_h} = n\mathbf{a_1} +
m\mathbf{a_2}$ and a translation vector $\mathbf{T_r} =
t_1\mathbf{a_1} + t_2\mathbf{a_2}$, where $(0 \le m \le n)$, $
\mathbf{a_1},\mathbf{a_2}$ are the lattice vectors, $t_1 = (2m +
n)/d_r$ and $t_2 = -(2n + m)/d_r$ with $d_r$ the greatest common
divisor of $(2n + m, 2m + n)$.

For electrons with group velocity $\mathbf{v}$, we define the
longitudinal component of the velocity to be ${\rm
v}_y=\mathbf{v}.\mathbf{T_r}$ and the transverse to be ${\rm
v}_x=\mathbf{v}.\mathbf{C_h}$. The quantized component $k_x$ of
the wave vector ${\bf{k}}=(k_x,k_y)$ parallel to the chiral vector
satisfies
\begin{displaymath}
     k_x^q=\frac{2\pi q}{L} ,(q=1,....N_{hex}),
\end{displaymath}
where $N_{hex}=2L^2/(a^2d_r)$ is the number of hexagons in the CNT
unit cell with $L$ the length of the chiral vector and $a$ the
length of the lattice vector. For a given value of the quantized
wavevector $k_x^q$, the energy of an electron is a function
$E(k_x^q,k_y)$ of the continuous, longitudinal component $k_y$ of
the wavevector, which satisfies
\begin{displaymath}
       -\frac{\pi}{\mathbf{|T_r|}}<k_ya<+\frac{\pi}{\mathbf{|T_r|}}
.\end{displaymath}

For example, within a minimal tight-binding model of $\pi$ bonding
in graphene, with a parameterized tight binding transfer integral
$\gamma$, the 1-D dispersion curves for such a CNT, are given by
\begin{eqnarray}
 E(k_x,k_y) = \pm \gamma (1+4\cos\frac{\sqrt{3}}{2}(k_xa\cos
 \theta^{}-k_ya
     \sin\theta^{})
      \nonumber\\
     \cos\frac{1}{2}(k_xa\sin \theta^{}+k_y a
     \cos\theta^{})+ \nonumber\\
      4\cos^{2}\frac{1}{2}(k_xa\sin \theta^{}+ k_ya \cos\theta^{}))^{1/2}
\end{eqnarray}
where $\tan\theta = (n-m)/{\sqrt 3}(n+m)$ (i.e. $\theta$ is
$\pi/6$ minus the chiral angle). For all electrons of energy $E$
moving along a 1-d channel of quantum number $q$, the equation
$E(k_x^q,k_y)=E$ can be solved to yield two $E$-dependent values
of $k_y$, which we denote by $k_y(q,E)$ and $-k_y(q',E)$,
corresponding to positive (ie. right-moving) and negative (ie.
left-moving) longitudinal group velocities ${\rm v}_y(q,k_y)$ and
${\rm v}_y(q',-k_y)$ respectively, where ${\rm
v}_y(q,k_y)=(1/\hbar)\partial E(k_x^q,k_y)/\partial k_y$. The
corresponding tangential group velocities are ${\rm v}_x(q,k_y)$
and ${\rm v}_x(q',-k_y)$ respectively, where ${\rm
v}_x(q,k_y)=(1/\hbar)\partial E(k_x^q,k_y)/\partial k_x^q$.

For an $(n',m')@(n,m)$ structure of the kind shown in
Fig.~\ref{Fig1}, which is derived from a chiral $(n,m)$ CNT inside
an achiral $(n',m')$ CNT, a right moving electron of energy $E$
incident along channel $q$, with tangential group velocity ${\rm
v}_x(q,k_y(q,E))$ will either be reflected into left-moving channels
$q'$ of the left CNT with tangential group velocities ${\rm
v}_x(q',-k_y(q',E))$ and reflection probability $R_{q'q}(E)$ or it
will be transmitted into right-moving channels $q'$ of the right CNT
with tangential group velocities ${\rm v}_x(q',k_y(q',E))$ and
transmission probability $T_{q'q}(E)$.

The flux of tangential momentum into the motor is
\begin{eqnarray}
  {\rm P_{ motor}}=\frac{2m_{e}}{h}
 \nonumber\\
  \int_{-\infty}^\infty {\rm dE
  [v_{in}}(E) -  {\rm v_{out}}(E)][f_{\rm{left}}(E) -
   f_{\rm{right}}(E)]
\end{eqnarray}
where $f_{\rm left}(E)$ and $f_{\rm right}(E)$ are Fermi
distributions for incoming channels from the left and right
reservoirs and
\begin{equation}
{\rm v_{in}}= \sum_q {{\rm v}}_x(q,k_y(q,E)).
\end{equation}
Furthermore,
\begin{equation}
{\rm v}\rm _{out}(E)={\rm v}_{\rm R}(E)+{\rm v}_{\rm T}(E),
\end{equation}
where
\begin{equation}
{\rm v_{R}}(E)= \sum_{q'q}[ {\rm v}_x(q',-k_y(q',E))R_{q'q}(E)]
\end{equation}
\begin{equation}
{\rm v_{T}}(E)=\sum_{q'q}[{\rm v}_x(q',k_y(q',E))T_{q'q}(E)]
\end{equation}
 In these equations, $q$ sums over incoming channels
and $q'$ sums over outgoing channels. Eq. (2) is valid even if
both CNTs are chiral. By symmetry, for the case considered here,
where the outer CNT is achiral,  ${\rm v_{\rm in}}(E)=0$. It can
also be demonstrated that for the case of total reflection, where
$ T_{q'q}(E)=0$ (for all $q',q$), equation (4) yields ${\rm
v_{out}}(E)=0$.

By Newton's third law, the net tangential force exerted on the inner
chiral CNT is ${\rm F_{motor} = - P_{motor}}$ and therefore
\begin{eqnarray}
{\rm F_{motor}}= \int_{-\infty}^{\infty} \frac{dE}{e} {\rm
f_{motor}}(E)[f_{\rm left}(E) - f_{\rm right}(E)]
\end{eqnarray} where ${\rm f_{motor}}(E)$ is the tangential
force per volt due to electrons of energy $E$, given by
\begin{equation}
{\rm f_{motor}}(E) =f_0[{\rm v}_{\rm out}(E)-{\rm v}_{\rm
in}(E)]/{\rm v}_{\rm F},
\end{equation}
with $f_0$ the characteristic tangential force per volt, given by
\begin{equation}
f_0 =\frac{2m_{e}e}{h}{\rm v}_{\rm F}.
\end{equation}

For metallic CNTs, ${\rm v}_{\rm F}= 9.35 \times 10^{5}$ m/s and
therefore $f_0= 4.12 \times 10^{-10}$N/V. (The corresponding force
at 1 volt is $F_0= 4.12 \times 10^{-10}$N and is the natural unit
of force introduced above.) We note that if $e\phi$ is the
chemical potential difference between the reservoirs and if
$e\phi$ is small compared with $k_BT$, then $[f_{\rm left}(E) -
f_{\rm right}(E)] \approx e\phi[-\partial f(E)/\partial E]$, which
reduces to $e\phi\delta (E-E_F)$ in the limit $k_BT \rightarrow
0$. On the other hand if $k_BT$ is small compared with $e\phi$,
then $[f_{\rm left}(E) - f_{\rm right}(E)]$ is of order unity
within an energy interval of order $e\phi$ and zero outside this
interval. In what follows, for the purpose of proving the
viability of the proposed motor, we shall focus on the latter
regime.

For a $(n',m')@(n,m)$ motor with an achiral $(n',m')$ outer CNT,
an upper bound for the tangential force is obtained by setting
$\rm v_{in}=0$ and replacing $\rm v_{out}$ by the maximum possible
tangential velocity carried by right-moving channels of an
infinite (n,m) chiral CNT. In units of ${\rm v_{F}}$ this is given
by
\begin{equation}
{\rm v_{\infty}}^{(n,m)}(E)= \sum_{q}{\rm v}_x^{(n,m)}(q,
k_y(q,E))/{\rm v_{F}},
\end{equation}
where ${\rm v}_x^{(n,m)}(q, k_y(q,E))/{\rm v_{F}}$ is the
dimensionless tangential group velocity of a right moving channel
$q$ of an infinite $(n,m)$ CNT.

The flux of tangential momentum carried by these electrons yields
an upper bound on the tangential force of the corresponding motor,
given by
\begin{equation}
F^{(n,m)}_{\infty}(e\phi)=\int^{e\phi/2}_{-e\phi/2} \frac{dE}{e}
{\rm f}^{(n,m)}_{\rm \infty}(E),
\end{equation}
where ${\rm f}^{(n,m)}_{\rm \infty}(E)$ is the flux of tangential
momentum per volt due to electrons of energy $E$, given by
\begin{equation}
{\rm f}^{(n,m)}_{\rm \infty}(E) =f_0{\rm v}_\infty^{(n,m)} (E).
\end{equation}

The latter quantity is shown as the red curve in Fig.~\ref{Fig2}b,
whereas the black curve shows the tangential force per volt,
obtained by computing all scattering coefficients and evaluating
equation (8).

To understand the origin of the oscillations in ${\rm
f}^{(n,m)}_{\rm \infty}(E)$, the red curves in Fig.~\ref{Fig2}c
show the values of ${\rm v}_x^{(6,4)}(q, k_y(q,E))/{\rm v_{F}}$
for right-moving channels. (for comparison, the blue curves show
 ${\rm v}_x^{(6,4)}(q', k_y(q',E))/{\rm v_{F}}$ for
left-moving channels). In equation (10), the label $q$ sums over
$N(E)$ open channels of the CNT, where $N(E)$ is an integer given
by $N(E) =\sum_{q}$. Clearly $N(E)$ is a discontinuous function of
$E$, which changes by an integer whenever new channels open or
close. As shown by the red curves in Fig.~\ref{Fig2}c,
right-moving channels open or close in pairs and just as a pair of
channels open, their tangential velocities cancel. Consequently,
${\rm v}_\infty^{(6,4)} (E)$ is a continuous function of $E$, with
a discontinuous first derivative.

The red curve of Fig.~\ref{Fig2}b shows that the tangential
velocity of right-moving electrons in an infinite chiral CNT is an
oscillatory function of $E$, whose slope changes whenever new
channels open. The black curve of Fig.~\ref{Fig2}b shows that
these oscillations are also present in ${\rm f_{motor}}(E)$. To
compute ${\rm f_{motor}}(E)$, we used a parameterized single-state
$\pi$ orbital Hamiltonian, based on a global fit to density
functional results for graphite, diamond and C$_2$ as a function
of the lattice parameter \cite{Tomanek1991}. This gives good
agreement with the single $\pi$ orbital interaction between two
carbon atoms separated by $d=3.40\AA$ predicted by
\cite{Saito1998} of 0.35 eV and is appropriate to our system,
where the inter-wall separation is ~$d=3.65\AA$. A recursive
Greens function formalism was then used to evaluate the
transmission matrix {\bf t}, describing the scattering of
electrons of energy $E$ from one end of the semi-infinite nanotube
to the other \cite{Sanvito1999}.

The tangential force shown in Fig.~\ref{Fig2}a is obtained by
integrating the black curve of Fig.~\ref{Fig2}b over the applied
voltage, in accordance with equation (7). The result depends
slightly on the  orientation of the inner tube relative to the outer
tube and therefore the results shown in Fig.~\ref{Fig2}a is averaged
over all orientations.  We have carried out calculations which
demonstrate the viabilities of CNT motors and drills with a wide
variety of chiralities and inter-tube couplings. The result is
rather robust and does not depend strongly on the number of overlap
atoms, provided the electrons evolving from the outer to the inner
tube do not strongly back scatter. If the intertube coupling causes
strong back scattering, then the tangential force is reduced, but
oscillations of the form shown in Fig.~\ref{Fig2}b persist.
Fig.~\ref{Fig2}b demonstrates that the tangential force vanishes for
energies below approx. 0.19 eV, since there are no open channels in
the chiral (6,4) CNT. It is interesting to note that a voltage
threshold also occurs when the achiral outer CNT is chosen to be an
armchair tube, because the two channels closest to the Fermi energy
possess vanishing tangential velocities and therefore electrons in
these channels cannot exert a tangential force. In all cases, the
voltage threshold can be removed by tuning the Fermi energy to
coincide with open channels of the outer CNT carrying a finite
tangential velocity.

In summary, we have highlighted a new force, arising from an
electron wind impinging on a chiral CNT, which is capable of
overcoming measured friction forces and driving a rotational CNT
motor or drill.  This may open a new route to low-power,
non-volatile nanoscale memory and could have applications to
nanofluidics. For the future, it will be of interest to examine
other drive mechanisms for chiral CNT motors. For example, if the
electrical contacts in Fig. 1 are replaced by reservoirs of atoms
or molecules and a pressure difference applied to drive the atoms
or molecules from left to right, then the resulting transfer of
angular momentum may also be sufficient to drive the motor, as
could a flux of phonons resulting from a temperature difference
between the ends of the device.

Finally, we note that other mechanisms for driving CNT motors have
been proposed, based on ac fields. In the case of
\cite{Kral2002}this involves the use of circularly polarized
light, whereas \cite{Tu2005} involves a brownian ratchet. The
force produced by the former is 2-3 orders of magnitude smaller
than the the drive mechanism discussed in the present paper, while
the latter requires a high-frequency drive voltage. By dispensing
with the need for electrostatic gates and ac voltages, fewer
processing steps are needed to construct the electron turbines of
Fig 1.

Acknowledgements.

This work is supported by the EPSRC, the EU MCRTN Fundamentals of
Nanoelectronics, the DTI, the RCUK Basic Technology programme and
North West Science Grid .

\end{document}